\documentclass[%
reprint,
superscriptaddress,
amsmath,
amssymb,
aps,
pra,
floatfix,
citeautoscript,
longbibliography
]{revtex4-1}

\usepackage{graphicx}
\usepackage{dcolumn}
\usepackage{bm}
\usepackage{units}
\usepackage{braket}
\usepackage{epstopdf}
\usepackage{color}
\usepackage{float}
\usepackage{csquotes}
\usepackage{array}

\usepackage[colorlinks, linkcolor=blue, citecolor=blue, urlcolor=blue, breaklinks=true]{hyperref}

\graphicspath{{./figures/}}
\usepackage[dvipsnames]{xcolor}
\usepackage[normalem]{ulem}

\begin{document}
\title{
Remotely Establishing Polarization Entanglement over Noisy Polarization Channels
}

\author{Sebastian Ecker}
\email{sebastian.ecker@oeaw.ac.at}
\affiliation{Institute for Quantum Optics and Quantum Information (IQOQI), Austrian Academy of Sciences, Boltzmanngasse 3, 1090 Vienna, Austria}
\affiliation{Vienna Center for Quantum Science and Technology (VCQ), Faculty of Physics, University of Vienna, Boltzmanngasse 5, 1090 Vienna, Austria}

\author{Philipp Sohr}
\affiliation{Institute for Quantum Optics and Quantum Information (IQOQI), Austrian Academy of Sciences, Boltzmanngasse 3, 1090 Vienna, Austria}
\affiliation{Vienna Center for Quantum Science and Technology (VCQ), Faculty of Physics, University of Vienna, Boltzmanngasse 5, 1090 Vienna, Austria}

\author{Lukas Bulla}
\affiliation{Institute for Quantum Optics and Quantum Information (IQOQI), Austrian Academy of Sciences, Boltzmanngasse 3, 1090 Vienna, Austria}
\affiliation{Vienna Center for Quantum Science and Technology (VCQ), Faculty of Physics, University of Vienna, Boltzmanngasse 5, 1090 Vienna, Austria}

\author{Rupert Ursin}
\email{rupert.ursin@oeaw.ac.at}
\affiliation{Institute for Quantum Optics and Quantum Information (IQOQI), Austrian Academy of Sciences, Boltzmanngasse 3, 1090 Vienna, Austria}
\affiliation{Vienna Center for Quantum Science and Technology (VCQ), Faculty of Physics, University of Vienna, Boltzmanngasse 5, 1090 Vienna, Austria}

\author{Martin Bohmann}
\email{martin.bohmann@oeaw.ac.at}
\affiliation{Institute for Quantum Optics and Quantum Information (IQOQI), Austrian Academy of Sciences, Boltzmanngasse 3, 1090 Vienna, Austria}
\affiliation{Vienna Center for Quantum Science and Technology (VCQ), Faculty of Physics, University of Vienna, Boltzmanngasse 5, 1090 Vienna, Austria}
\affiliation{Department of Experimental Physics, Comenius University, SK-84248 Bratislava, Slovakia}

\begin{abstract}

The faithful distribution of entanglement over noisy channels is a vital prerequisite for many quantum technological applications. 
Quantum information can be encoded in different degrees of freedom (DoF) of photons, where each encoding comes with its own advantages and disadvantages with respect to noise resilience and practicality in manipulation. 
In this work, we experimentally implement a deterministic entanglement purification protocol that allows us to faithfully distribute entanglement in one DoF over a noisy channel and then remotely transfer it to another DoF for manipulation.
In particular, we distribute robust energy-time entanglement and transfer it to polarization entanglement at the communicating parties. 
The remotely obtained polarization state is independent of the polarization noise during distribution and reaches fidelities to a Bell state of up to $97.6\%$.
Our scheme enables robust and efficient polarization entanglement distribution in the presence of arbitrary polarization noise, which is relevant for future large-scale quantum networks.
\end{abstract}

\maketitle

\section{Introduction}
Entanglement is arguably the most important resource for quantum information processing and its distribution between remote parties is one of the key challenges in quantum technology. 
The uncontested carriers of quantum information are photons due to their speed, relative robustness towards decoherence, and high generation rates.
Entangled pairs of photons are therefore an indispensable resource for entanglement distribution over long-distance fiber \cite{Wengerowsky19, marcikic04} and free-space \cite{ecker2021,yin17} links. 
Various photonic degrees of freedom (DoF) can be harnessed for quantum information encoding. 
Among others, two-photon entanglement has been demonstrated in the polarization \cite{kwiat95,Wengerowsky19,ecker2021,yin17}, frequency \cite{yuan20,olislager10,xie2015,lu2018,bernhard13,kues2017}, temporal  \cite{raymer2020,brecht2015,martin17}, path \cite{rarity90,schaeff2012,lee17} and transverse spatial \cite{mair2001,erhard2018} DoF. 

\begin{figure*}[t]
\centering
\includegraphics[width=1\textwidth]{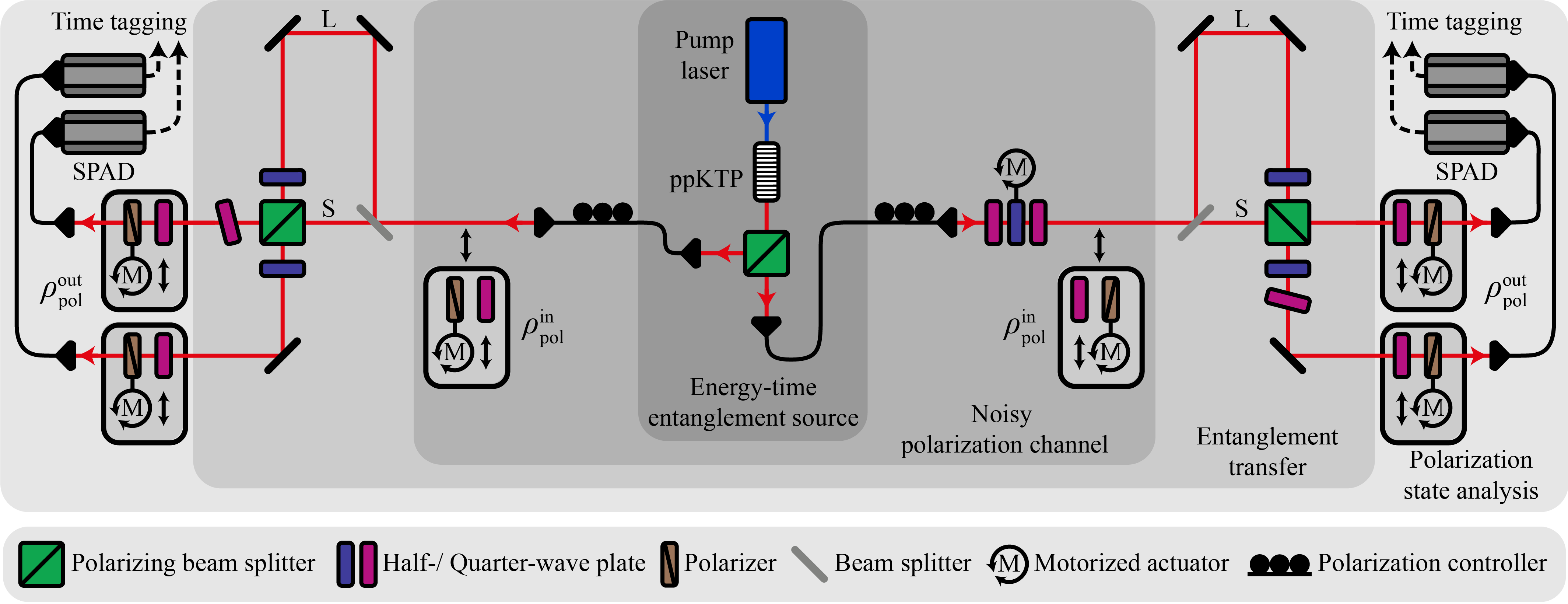}
\caption{
Experimental setup. Energy-time entangled photon pairs are produced by focusing a spectrally narrowband pump laser into a periodically poled potassium titanyl phosphate (ppKTP) crystal. Spontaneously created photons are then coupled into single-mode fibers and guided to the communicating parties via a noisy polarization channel. After the channel, the entanglement is transferred from the energy-time domain to the polarization domain using a Franson-type interferometer, which is actively phase-stabilized. A polarization state tomography setup can be inserted before and after the transfer (gray boxes). The photons are registered by single-photon avalanche diodes (SPADs), time-tagged, and recorded for postprocessing.}
\label{fig:setup}
\end{figure*}

Polarization-entangled photon pairs are routinely used in entanglement distribution \cite{yin17,Wengerowsky19}.
While the polarization state of photons can be efficiently manipulated, it is susceptible to noise influences outside of protected laboratory environments.
The resulting errors are mainly caused by stress- and temperature-induced birefringence \cite{Ulrich80,Ren88} and polarization-mode dispersion \cite{Gisin91} over long optical fiber links. 
Another source of errors in the polarization domain is reference-frame misalignment between the communicating parties \cite{souza08,laing10,dambrosio12, Tannous2019}.
The latter will become increasingly relevant for large-scale quantum networks consisting of reconfigurable quantum links between satellites \cite{boone15}, drones \cite{Liu21}, and ground stations \cite{yin17}.
While the transverse spatial and path DoF of photons can also be efficiently manipulated, they face even higher disturbances outside of controlled laboratory settings, necessitating characterization of static channels \cite{valencia2020}, active phase stabilization over the channel \cite{dalio20}, or incoherent detection schemes \cite{krenn15}.

One possibility of circumventing error-prone entanglement distribution is the exploitation of the temporal and frequency DoF. 
These intrinsic DoF do not depend on a shared spatial reference frame between the communicating parties and are inherently error resilient during transmission \cite{brecht2015,maclean18}.
On the downside, the manipulation and measurement of quantum states encoded in these DoF is elaborate and involve optical modulators or nonlinear detection schemes \cite{Lukens:17,brecht2015,Raymer20time}, which makes their utilization in quantum processing challenging. 
Therefore, frequency and temporal-mode entanglement is best suited for entanglement distribution, while the polarization or spatial domain qualifies for state manipulation and detection.
This requires an efficient entanglement transfer between these DoF in order to simultaneously achieve robust entanglement distribution and efficient state manipulation. 

A good candidate for implementing this transfer is deterministic entanglement purification \cite{li2010,sheng10onestep,sheng10deterministic}.
In contrast to entanglement distillation, which uses several entangled photon pairs \cite{bennett96,deutsch96} or different DoF \cite{ecker2021experimental, hu21} of a single photon pair for increasing the entanglement probabilistically, this approach can be used to deterministically transfer entanglement from one DoF to another DoF.
Importantly, the quality of the entangled state after the transfer is independent of the state in the same DoF prior to the transfer \cite{li2010,sheng10onestep}. 
Hence, deterministic entanglement purification protocols provide a promising pathway to noise-resilient entanglement distribution. 

In this contribution, we utilize energy-time entangled photon pairs produced in spontaneous parametric down-conversion (SPDC) for robust entanglement distribution and subsequently implement a deterministic entanglement purification protocol by means of a Franson-type interferometer to transfer the entanglement to the polarization DoF.
This approach allows us to establish polarization entanglement over a noisy polarization channel with unit protocol efficiency.
The resulting polarization-entangled state exhibits fidelities to a Bell state of up to \unit[97.6]{\%}, which is comparable to that of state-of-the-art polarization-entangled photon-pair sources. 
By combining the resilience of the energy-time domain for long-distance entanglement distribution with the versatility of polarization states for efficient manipulation and detection, we exploit the best of both worlds.

\section{Experiment}
The experimental setup consists of an entangled photon-pair source, a noisy polarization channel, and an entanglement transfer stage followed by a polarization state measurement (see Fig.~\ref{fig:setup}).
Energy conservation in the SPDC process and temporal coherence of the pump field lead to the emission of photon pairs entangled in the energy-time DoF \cite{kwiat93}. 
Our continuous-wave pump laser has a spectral bandwidth of $\Delta \nu_\text{FWHM} < \unit[1]{MHz}$ at a wavelength of \unit[404.53]{nm}. 
For long-term phase stability, the laser is frequency locked to a saturated absorption spectrum of ${}^{39}\text{K}$, resulting in a wavelength stability of $\sim\unit[0.6]{fm/min}$. 
The pump field spontaneously creates photon pairs in a $30\,$-mm-long periodically poled potassium titanyl phosphate (ppKTP) crystal, temperature tuned for degenerate type-II phase matching. 
We utilize the orthogonal polarization of the collinearly propagating photon pairs for their separation and subsequently couple them into single-mode fibers. 

The energy-time entangled photon pairs are then distributed to the communicating parties through a noisy polarization channel, simulating a noisy environment. 
The polarization state is coherently manipulated by in-fiber polarization controllers and incoherently manipulated by time averaging over a rotating wave plate.
Averaging over one revolution of the wave plate, a pure single-photon input state $\ket{\text{H}}$ is transformed into a maximally mixed state $(\ket{\text{H}}\bra{\text{H}} + \ket{\text{V}}\bra{\text{V}})/2$, where H (V) is the horizontal (vertical) polarization component.
Hence, we realize mixed states via ensemble averaging~\cite{Bohmann2017,Cimini2019}.

We accomplish entanglement transfer from the energy-time domain to the polarization domain by a Franson-type interferometer \cite{franson89}.
The imbalanced Mach-Zehnder interferometers probabilistically convert the temporal modes $\ket{t}$ and $\ket{t+\Delta t}$ into interferometer path modes $\ket{L}$ and $\ket{S}$.
The difference in the path lengths of the long (L) and short (S) arms of the interferometers corresponds to a temporal imbalance of $\Delta t = \unit[2.6]{ns}$.
Subsequently, the paths are interfered on polarizing beam splitters, which act as two-qubit controlled NOT (CNOT) gates between the path and the polarization DoF \cite{ecker2021experimental}. 
Here, the polarization state of the photon (control qubit) determines the state of the output path modes of the polarizing beam splitter (target qubit).
These two CNOT gates, along with half-wave plates for bit-flip operations, constitute a deterministic entanglement purification protocol \cite{li2010}.
The desired entangled polarization state between all output ports is obtained by compensating for phase shifts introduced by the beam splitters with quarter-wave plates and by locking the interferometers to specific phase settings.
We stabilize the phase between the long and the short arms of the interferometers by injecting a portion of the frequency-locked pump laser into the interferometers and utilize the interference contrast as feedback for a phase actuator control loop~\cite{ecker2021experimental}.
After single-mode coupling, the energy-time entangled photon pairs exhibit a Franson visibility \cite{kwiat93} of \unit[97.9]{\%}.

In order to characterize the initial polarization state after the noisy channel $\rho_\text{pol}^\text{in}$ and the state after the entanglement transfer $\rho_\text{pol}^\text{out}$, we utilize quantum state tomography \cite{altepeter2005}. 
The tomography setup consists of a motorized linear polarizer and a removable quarter-wave plate, which enables projective measurements on arbitrary polarization states. 
For the tomography of $\rho_\text{pol}^\text{in}$, the polarizers are temporarily inserted and the long arms of the interferometers are blocked to inhibit interference. 
The photons are detected by single-photon avalanche diodes (SPAD), time-tagged, and recorded for postprocessing~\cite{ecker2021experimental}.
A photon pair is identified if both communicating parties register a detection within a coincidence window of 1 ns.
Depending on the initial state $\rho_\text{pol}^\text{in}$, photon-pair events are registered in four possible detector combinations.
Since the protocol is deterministic, none of these detector combinations are discarded, except for the usual coincidence postselection necessary for Franson interference~\cite{kwiat93}, and the bare protocol efficiency is thus 100 \%; see also Ref.~\cite{sheng10onestep}.
The integration time for a single measurement is \unit[25]{s}, which matches the rotation period of the noisy polarization channel wave plate. 
Typical single-photon rates for Alice and Bob are \unit[400]{kcps}, while the total coincidence rate between all detectors amounts to \unit[10.3]{kcps}. 

\section{Results}
The purification procedure interchanges the quantum states of the polarization and energy-time DoF.
In particular, an initially entangled state in the energy-time DoF is transferred to the polarization DoF: 
\begin{equation}
     \rho_\text{pol}^\mathrm{in} \otimes \boxed{\rho_\text{e-t}^\mathrm{in} \xrightarrow[\text{Transfer}]{\text{Entanglement}}
     \rho_\text{pol}^\mathrm{out}} \otimes \rho_\text{e-t}^\mathrm{out}.
     \label{eq:transfer}
 \end{equation}
Provided that the energy-time DoF is noiselessly distributed and our setup accesses a two-dimensional discretization in the form of a $\Phi^+$ Bell state
\begin{align}
\begin{split}
    \rho_\text{e-t}^\text{in} &=
    \ket{\Phi^+_\text{e-t}}\bra{\Phi^+_\text{e-t}} \\&=
    \frac{1}{2}(\mathrm{\ket{L,L}+\ket{S,S})}(\mathrm{\bra{L,L}+\bra{S,S})},
    \label{eq:Franson}
\end{split}
\end{align}
the purification operation transfers this state into the polarization domain, resulting in the state 
\begin{align}
\begin{split}
    \rho_\text{pol}^\text{out} &=
    \ket{\Phi^+_\text{pol}}\bra{\Phi^+_\text{pol}}\\&=\frac{1}{2}(\mathrm{\ket{H,H}+\ket{V,V}})(\mathrm{\bra{H,H}+\bra{V,V}}).
\end{split}
\end{align}
Ideally, the state $\rho_\text{pol}^\text{out}$ only depends on the quality of $\rho_\text{e-t}^\mathrm{in}$, while it is independent of the polarization state prior to the transfer $\rho_\text{pol}^\mathrm{in}$~\cite{sheng10onestep}.
This is critical for the success of the entanglement transfer, since the polarization state is neither preserved nor necessarily known after passing the noisy quantum channel.

In order to experimentally demonstrate this property, we consecutively prepare two orthogonal mixed states:
\begin{align}
\mathrm{\rho_{pol}^{in,1}} &= \frac{1}{2}\mathrm{(\ket{H,H}\bra{H,H}+\ket{V,H}\bra{V,H})} \ \mathrm{and} \label{eq:rhoin1}\\
\mathrm{\rho_{pol}^{in,2}} &= \mathrm{\frac{1}{2}(\ket{V,V}\bra{V,V}+\ket{H,V}\bra{H,V})}.\label{eq:rhoin2}
\end{align}
They are produced by incoherently manipulating one of the photons of the pure polarization states $\mathrm{\ket{H,V}}$ and $\mathrm{\ket{V,H}}$, respectively.
After their preparation, the states are used as input for the entanglement transfer, resulting in the purified states $\mathrm{\rho_{pol}^{out,1}}$ and $\mathrm{\rho_{pol}^{out,2}}$, respectively.
\begin{figure}
    \centering
    \includegraphics[width=1\columnwidth]{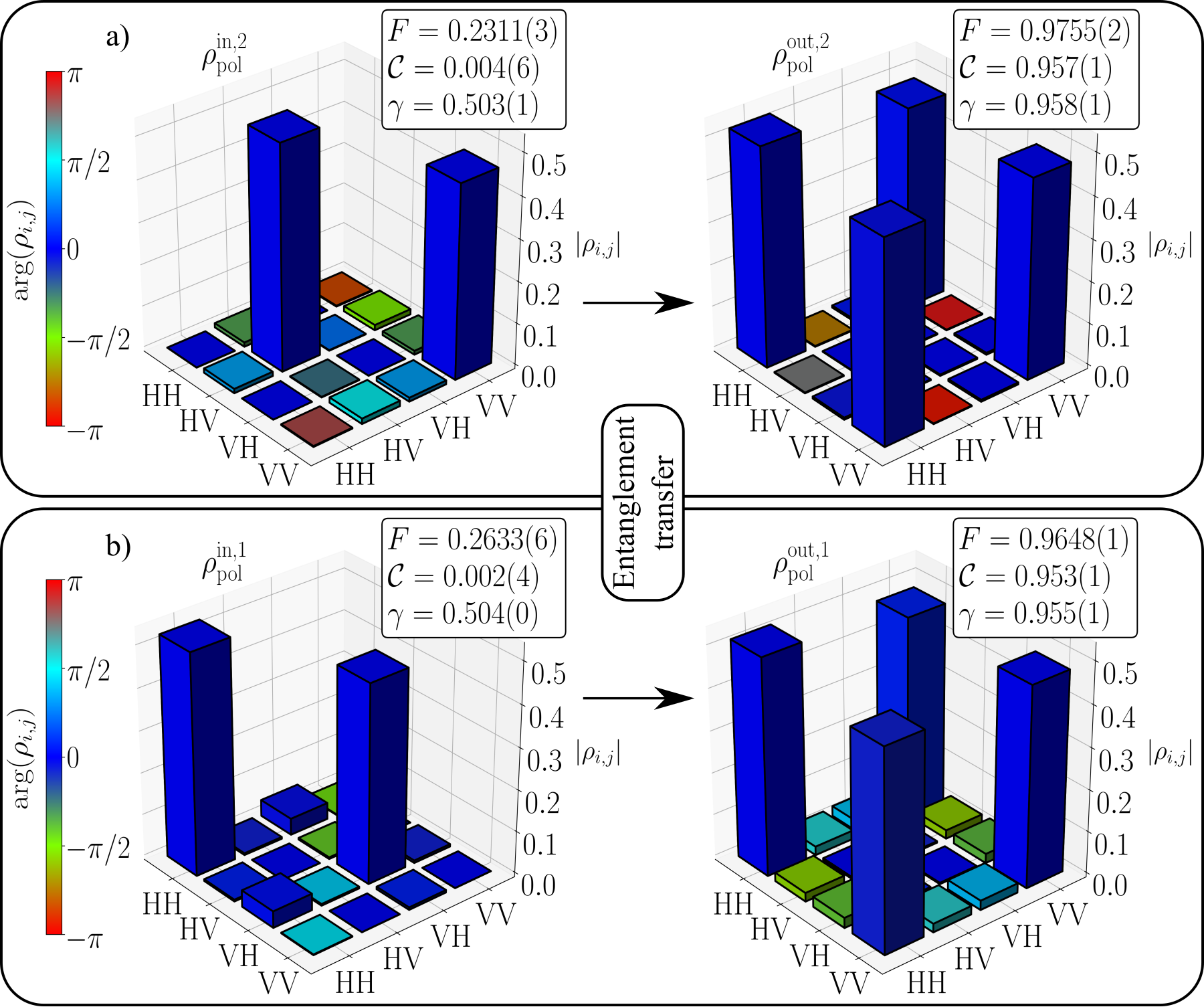}
    \caption{
    Reconstructed polarization density matrices before ($\rho_\mathrm{pol}^\mathrm{in}$) and after ($\rho_\mathrm{pol}^\mathrm{out}$) the entanglement transfer. a) Transfer from state $\rho_\mathrm{pol}^\mathrm{in,2}$ to $\rho_\mathrm{pol}^\mathrm{out,2}$. b) Transfer from state $\rho_\mathrm{pol}^\mathrm{in,1}$ to $\rho_\mathrm{pol}^\mathrm{out,1}$. The height of the bars corresponds to the magnitude $|\rho_{i,j}|$ of the density matrix elements $\rho_{i,j}$, and the color corresponds to their phase $\mathrm{arg}(\rho_{i,j})$. For each state, we calculate the fidelity to the $\Phi^+$ Bell state $F$, the concurrence $\mathcal{C}$, and the purity $\gamma$. The uncertainties of these quantities correspond to one standard deviation of the mean, calculated by propagating the Poissonian error in the count rates via a Monte Carlo simulation.}
    \label{fig:tomography}
\end{figure}
We perform a quantum state tomography on the state before and after the transfer and reconstruct the density matrices as illustrated in Fig.~\ref{fig:tomography}.
For each state we calculate the concurrence $\mathcal{C}$ \cite{Horodecki09}, the purity $\gamma = \text{tr}(\rho_\mathrm{pol}^2)$, and the fidelity to the $\Phi^+$ Bell state ${F=\bra{\Phi^+}\mathrm{\rho_{pol}}\ket{\Phi^+}}$.  
While the concurrence prior to the transfer is virtually zero, which is evidence of a separable state, it increases to almost 1 after the transfer, which implies that they are close-to-maximally entangled. This is further supported by the high values for $\gamma$ and $F$; cf. Fig. \ref{fig:tomography}.
Thus, our results show that we successfully created close-to-perfect polarization entanglement over a noisy polarization channel in a nonlocal fashion by means of a deterministic entanglement transfer between two DoF.
Furthermore, we also demonstrate that the underlying bilateral CNOT operations can purify arbitrary mixed polarization states with near unit efficiency.

In a second analysis, we investigate the entanglement transfer with nonmaximally entangled polarization input states.
To this end, we prepare a family of polarization states close to 
\begin{equation}
    \ket{\Phi^+_p} = \sqrt{p}\ket{\mathrm{H,H}} + \sqrt{1-p}\ket{\mathrm{V,V}},
\label{eq:p_state}
\end{equation}
with $p\in[0,0.5]$, which are parametrized from separable ($p=0$) to maximally entangled ($p=0.5$).
The states are created by placing the SPDC crystal of our setup in the center of a Sagnac interferometer \cite{fedrizzi07,kim06, ecker2021experimental}. 
In this configuration, the balance parameter $p$ of the state can be tuned by adjusting the power ratio of the pump laser between the two interferometer directions by means of the pump laser polarization. 
For a balanced power splitting ($p=0.5$), the source produces a maximally hyperentangled state in the energy-time and polarization domains \cite{Kwiat97,steinlechner17, ecker2021experimental}
The experimental balance parameter $p_\text{exp}$ is determined from the number of registered photons in state $\ket{\mathrm{H,H}}$ and $\ket{\mathrm{V,V}}$, respectively.

\begin{figure}
        \centering
        \includegraphics[width=1\columnwidth]{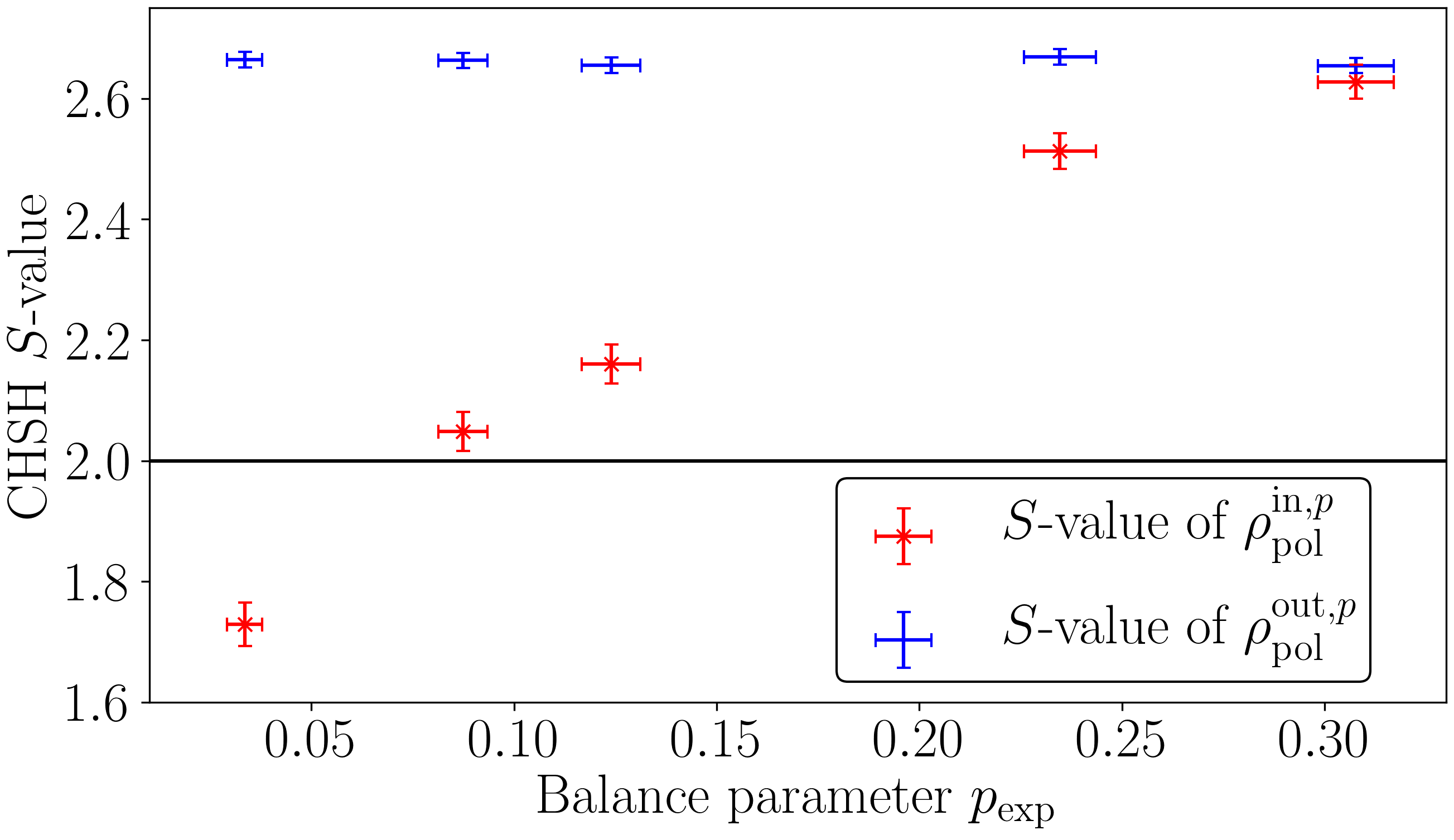}
        \caption{Bell test for various nonmaximally entangled input states of the form in Eq. \eqref{eq:p_state}. For each balance parameter $p$ of the state, the CHSH $S$-value is evaluated before ($\rho_\mathrm{pol}^{\mathrm{in},p}$) and after ($\rho_\mathrm{pol}^{\mathrm{out},p}$) the entanglement transfer.  $|\langle\hat{S}\rangle|>2$ certifies entanglement and Bell nonlocality. The error bars in both directions correspond to 5$\sigma$ standard deviations due to the count statistics.}
        \label{fig:CHSH}
\end{figure}

In order to study and certify Bell nonlocality \cite{Brunner14} and entanglement of the produced states, we employ the Clauser-Horne-Shimony-Holt (CHSH) $S$-value \cite{clauser69}:
\begin{equation}
    \langle\hat{S}\rangle = \langle\hat{\sigma}_\alpha\hat{\sigma}_\beta\rangle-\langle\hat{\sigma}_\alpha\hat{\sigma}_{\beta'}\rangle+\langle\hat{\sigma}_{\alpha'}\hat{\sigma}_\beta\rangle+\langle\hat{\sigma}_{\alpha'}\hat{\sigma}_{\beta'}\rangle .   
\end{equation}
Here, the operators $\hat{\sigma}$ are Pauli operators with linear polarization eigenstates characterized by the angles $\alpha = 0^{\circ}$, $\alpha'= 45^{\circ}$ and $\beta=22.5^{\circ}$, $\beta'=67.5^{\circ}$, respectively (see, e.g., Ref.~\cite{Brunner14}).
A value of $|\langle\hat{S}\rangle|>2$ is incompatible with local hidden variable theories and provides an entanglement-witness condition. 
We measure the $S$-value \cite{noteLoophole} for different states parametrized by the balance parameter $p$ before ($\rho_\mathrm{pol}^{\mathrm{in},p}$) and after ($\rho_\mathrm{pol}^{\mathrm{out},p}$) the entanglement transfer.
Our results show that the state after the transfer $\rho_\mathrm{pol}^{\mathrm{out},p}$ is close-to-maximally entangled irrespective of the balance parameter (see Fig.~\ref{fig:CHSH}).
Hence, the implemented scheme can also be used to increase the entanglement of nonmaximally entangled states and can even enable a violation of a Bell-type inequality in the polarization DoF.

\section{Discussion}
We demonstrated an efficient method for remotely preparing high-fidelity polarization entanglement despite prior distribution of the entangled photon pairs via noisy polarization channels.
This is enabled by utilizing entanglement in the energy-time domain of photon pairs, which is robust during distribution.
After distribution, the entangled state is deterministically transferred to the polarization DoF by means of a deterministic entanglement purification protocol. 
We experimentally tested the transfer by consecutively feeding it with mixed states passing a tunable noisy polarization channel and nonmaximally entangled states. 
For all generated states, high-fidelity polarization entanglement was retrieved by the entanglement transfer.
In another recent experimental realization of deterministic entanglement purification~\cite{Huang2021}, the spatial and polarization domain of photons was utilized to demonstrate a slightly different protocol~\cite{Sheng2010}.

Polarization entanglement is conventionally created by locally superposing two-photon states created in SPDC \cite{Anwar21}.
We, on the other hand, produced polarization entanglement remotely at the receiving parties after the noisy polarization channel and achieved polarization-state fidelities that can compete with state-of-the-art polarization-entangled photon-pair sources \cite{chen18,Anwar21}. 
In contrast to polarization entanglement, which requires elaborate source engineering to achieve spatiotemporal mode overlap \cite{Anwar21}, energy-time entanglement arises from the SPDC process itself. 
We are therefore harnessing entanglement that is created and distributed as a byproduct of many long-distance entanglement-distribution experiments, where its potential is commonly unexploited \cite{Wengerowsky19,yin17}. 
While interference after multimode channels such as free-space links is hindered by wave-front distortions, several experiments tackled this issue in an efficient manner~\cite{jin18,Jin:19}.
The entanglement transfer shifts the creation of polarization entanglement from the source to the receiving parties.
Unlike in common polarization entanglement sources, the entanglement fidelity of our scheme is limited by spatiotemporal mode overlap in the Franson interferometer.
This shift of complexity is particularly relevant for the long-distance distribution of entanglement via satellite down-links \cite{yin17}, where the harsh conditions on a satellite during launch and in-orbit operation are extremely challenging for the interferometric stability of the source. 
Furthermore, the entanglement transfer allows one to remotely establish polarization entanglement without the need of a shared spatial reference frame.
This makes active polarization control redundant and provides an elegant solution for reference-frame-independent quantum communication that otherwise requires the sending of multiple photons \cite{chen06} or the faithful transmitting of transverse spatial modes \cite{aolita07, souza08, dambrosio12,laing10}. 
While this is the case for fiber-based entanglement distribution without caveat, for fast-moving satellites in a low Earth orbit another problem arises.
Due to the relativistic Doppler shift~\cite{einstein1905}, the phase of the entangled energy-time state shared between satellite and ground station is shifted, thus adversely affecting the polarization state.
This frequency shift can be compensated by adjusting one of the interferometer imbalances accordingly~\cite{chapman2020}.

It is important to discuss the relation between the implemented deterministic entanglement purification protocol and entanglement distillation.
In the latter case, the output state after one iteration of the protocol is closer to the maximally entangled state than the input state \cite{bennett96,deutsch96}.
However, entanglement distillation has a nonunit protocol efficiency and the maximally entangled state can only be approached asymptotically after many distillation steps.
On the other hand, in deterministic entanglement purification, the quantum state of one DoF is merely transferred to another DoF. 
Hence, while the implemented protocol does not perform a distillation of entanglement, it provides a deterministic way to obtain a maximally entangled state in a single step, provided that the auxiliary DoF is maximally entangled.
Notably, this state transfer works for any input state and also maps the input polarization state on the path modes after the interferometer.
This general working principle can be further exploited in quantum information applications that rely on the transfer of information between different DoF.
For example, it could be used to transfer quantum information to the polarization DoF in order store it in quantum memories that are interfaced with the polarization DoF \cite{specht2011,lvovsky2009}, thus facilitating quantum repeater networks~\cite{briegel98,duer99}.

There are several possible extensions of our experiment.
Firstly, the entanglement transfer could be adapted to other photonic DoF with appropriate CNOT gates, broadening the scope of our approach.
Secondly, the energy-time DoF of photon pairs generated in SPDC is intrinsically high-dimensionally entangled \cite{khan06}, which opens up the possibility for entanglement purification of high-dimensional states~\cite{MiguelRamiro2018}. An extension to deterministic purification protocols would allow one to combine the benefits of high-dimensional entanglement, such as resilience to noise \cite{ecker19} and increased channel capacity \cite{barreiro08}, with the versatility provided by the entanglement transfer. 
Altogether, we have demonstrated a method for remotely establishing entanglement over noisy channels, opening up many applications for practical quantum information technologies. 

We thank Marcus Huber for fruitful discussions. 
We acknowledge the European Union’s Horizon 2020 programme grant agreement No. 857156 (OpenQKD) and the Austrian Academy of Sciences in cooperation with the FhG ICON-Programm “Integrated  Photonic  Solutions for Quantum Technologies” (InteQuant). We also gratefully acknowledge financial support from the Austrian Research Promotion Agency (FFG) Agentur für Luft- und Raumfahrt (FFG-ALR  contract  844360  and854022).


%

\end{document}